# Ballistic Supercavitating Nano Swimmer Driven by Single Gaussian Beam Optical Pushing and Pulling Forces

Eungkyu Lee[a], Dezhao Huang[a] and Tengfei Luo[a,b,c*]

[a]Department of Aerospace and Mechanical Engineering, University of Notre Dame, Notre Dame, USA

[b]Department of Chemical and Biomolecular Engineering, University of Notre Dame, Notre Dame, USA.

[c]Center for Sustainable Energy of Notre Dame (ND Energy), University of Notre Dame, Notre Dame, USA.

*Correspondence to: tluo@nd.edu

**Abstract**

Directed high-speed motion of nanoscale objects in fluids (nano swimmers) can have a wide range of applications like molecular machinery, nano robotics, drug delivery, and material assembly. Here, we report ballistic plasmonic Au nanoparticle (NP) swimmers with unprecedented speeds (~336,000 μm s$^{-1}$) realized by not only optical pushing but also pulling forces from a single Gaussian laser beam. Both the optical pulling and high swimmer speeds are made possible by a unique NP-laser interaction. The Au NP excited by the laser at the surface plasmon resonance peak can generate a nanoscale bubble, which can encapsulate the NP (*i.e.*, supercavitation) to create a virtually frictionless environment for it to move, like the Leidenfrost effect. While optical forces are mostly positive, certain NP-bubble configurations can lead to "negative" optical forces that pull the NP to swims against the photon stream. The demonstrated ultra-fast, light-driven NP movement may benefit a wide range of nano- and bio-applications and provide new insights to the field of "negative" optical force.

**Introduction**

Speed is important to the functionality of nano/micro swimmers in fluids because it determines the efficiency of swimmers[1-4]. Directed movement is also important as the swimmers are usually tasked to reach a target. However, directed movement and high speed in the nanoscale are rarely compatible[5-7]. In laminar fluid flow, enabling fast-moving swimmers requires large propulsion force, as it needs to counter the drag force which is proportional to the speed[8]. Usually, nano/micro swimmers driven by on-board propulsion (*e.g.*, bubble repellence[5-7]) exhibit higher speeds compared to other mechanisms (*e.g.*, mechanical motion[9,10], pressure[11] and thermal[12,13] gradients, chemical phoresies[14-16] and optical force[17,18]). The highest speed of bubble-repelled swimmers is ~$10^5$ body-length s$^{-1}$ (ref [7]), where the unit (the speed to body-length ratio) is commonly used to allow a fair comparison between swimmers with different dimensions. However, it is very difficult to control their moving directions because of the random nature of the on-board repellence force.

Optical forces on an object due to the exchange of photon momentum with the object[17-27] can move nano/micro swimmers along the beam direction. Light propulsion has unique advantages like wireless control, high spatial and temporal precision and instant response[28]. Usually, light applies a pushing force on an object in the light propagating direction, which has worked together with the optical gradient force to enable the optical tweezer effect[17,19,20]. Objects with high scattering efficiency like plasmonic NPs can enable stronger optical pushing forces[19,20], but the achieved highest swimmer speed is merely ~$10^3$ body-length s$^{-1}$ (ref [17]), since the optical forces are very weak ($10^{-12}$-$10^{-14}$ N).

Light may also pull an object in specific optical conditions[18,21-27]. For example, two planewaves irradiating mirror-symmetrically with a large incidence angle towards an object can lead to a strongly focused forward photon scattering momentum, resulting in a net "negative" optical force [22-24]. For a spheroidal object much larger than the wavelength of the incident light and placed at a heterogenous dielectric interface, the object can be pulled by the light due to an increase in the photon linear momentum across the dielectric interface[18]. However, the achieved moving speed is only ~ 0.5 body-length s$^{-1}$, and the same strategy may not be applicable to objects in homogenous media, which are the cases for most nano/micro swimmers. Theoretically, an object in homogenous medium can also be pulled by a single planewave if the object has certain unique optical configurations to enable either optical gain[25,26] or near-field electromagnetic coupling between dielectric-metallic NP dimers[27]. However, the experimental demonstration in homogenous media is still lacking. In this work, we observe optical pushing and pulling of Au NP swimmers exhibiting extremely high speeds (>$10^6$ body-length s$^{-1}$) in water, and we show that the unusual optical pulling and high-speed movement are only made possible by the nano-bubbles generated around the laser-excited plasmonic NPs, which lead to supercavitation.

**Results and Discussion**

The experimental setup is shown in Fig. 1a. We disperse Au core-shell NPs made of a silica core (~100 nm) and an Au shell (~10 nm) in deionized water. A femtosecond pulsed laser (linear-polarized Gaussian beam) with a repetition rate of 80.7 MHz and a wavelength of 800 nm is focused in the NP-water suspension using a 20× objective lens (numerical aperture ~0.42). At the focal plane, the minimum waist of the Gaussian beam is ~6 μm. The laser wavelength matches the surface plasmon resonance (SPR) peak of the Au NPs. A high-speed camera captures the scattered light from the Au NPs to track their positions, and they are shown as glowing dots (Figs. 1b and 1c).

By observing the glowing dots, we find some fast NPs moving along the beam propagation direction ("positive motion") and some in the opposite direction ("negative motion"). NPs moving in both directions can have very high speeds, in stark contrast to the rest NPs experiencing Brownian motion. We single out two NPs, each moving in one direction, and highlight their positions at different times in Figs. 1b and 1c. The highest speed for the positive motion is 336,000 μm s$^{-1}$ and that for the negative motion is 245,000 μm s$^{-1}$. We also find that the average speed of the positive motion (204,000 μm s$^{-1}$) is higher than that of the negative motion (112,000 μm s$^{-1}$).

When the Gaussian beam illuminates the NP, there are mainly three types of possible forces, namely the optical force (radiative pressure)[17,19,20], the optical gradient force[17,19,20] and the photothermal gradient force[12,13,29]. Gravity and buoyancy are not considered because they are perpendicular to the optical axis. To identify the dominant force driving the ballistic NPs, we analyze several representative ballistic trajectories around the focal plane as shown in Figs. 1d and 1e. It is found that the NPs can cross the focal plane (black arrow in Figs. 1d and 1e) regardless of their moving directions. In addition, we can see that to the left (or right) side of the focal plane, ballistic NPs can move in both directions. These indicate that the ballistic NPs are not driven by any gradient (optical or photothermal) forces, since such forces are symmetric about the focal plane, which should converge the NPs from both sides of the focal plane towards it. The optical gradient force is also indeed very small with the Gaussian beam used in our experiments which has a minimum waist of ~ 6 μm. We also find the same ballistic NPs in suspensions with different NPs concentrations, suggesting that the ballistic movements should not be the result of the environment (*e.g.*, scattered light from surrounding NPs or thermalization of water due to NP heating). These analyses leave the optical force (radiative pressure) as the only possible driver of the ballistic NPs.

To fully explain the observed ballistic NP movements, we need to understand how the NPs can move in speeds much higher than that a typical optical force can sustain according to Stokes' law and why there are both optical pushing and pulling phenomena on some specific NPs. As discussed later, the optical force from our laser cannot sustain the observed high speed unless the effective viscosity experienced by the NPs is 100× lower than that of liquid water. The optical pulling is also impossible with our Gaussian beam irradiating an Au NP in a homogenous medium, unless there is a nearby dielectric structure optically coupled to the NP like the ones proposed in Refs.[21,27].

According to several studies[30-33], when an Au NP is irradiated by a laser at the SPR peak, it can instantaneously generate a nano-bubble around it. Using a pump-probe optical scattering imaging setup, we confirmed the existence of nano-bubbles around the plasmonic Au NPs in our experiments when excited by the pump laser (Fig. 1f and 1g). This pump-probe method detects the intensity change of the scattered probe light (533 nm) at a certain solid angle due to the formation of nano-bubble on Au NP under the illumination of the pump beam at the SPR peak (800 nm). We note that the lifetimes of the plasmonic nanobubbles are reported to be ~200 ns (Refs.[30-33]). In our NP movement experiments, the time interval between laser pulses is ~12 ns, allowing the nano-bubbles to shrink due to cooling by the surrounding water[30,31,33-35], but they may not completely collapse since the pulse period (~12 ns) is shorter than the bubble lifetime (~200 ns). Dissolved gas in water can also contribute to the nano-bubble volume, making it more stable than a pure vapor bubble[36].

To understand how the nano-bubbles may influence the optical force direction, we calculate the optical force in the beam propagation direction ($F_z$) on an NP with a nano-bubble. We consider a bubble with a radius $r_{nb}$ < 120 nm attached to the surface ($\theta,\varphi$) of an Au core-shell NP ($\theta$ and $\varphi$ are polar and azimuthal angles, Fig. 2a). When the bubble keeps growing, it will eventually encapsulate the NP, and we assume such a case for $r_{nb}$ > 120 nm (*i.e.*, inset in Fig. 2b). We note that the exact bubble radius for the attach-to-encapsulate transition will not influence the physics inferred from our analysis. Since the beam waist (~6 μm) at the focal plane is much larger than the NP size (~120 nm), and the ballistic movements start at locations away from the focal plane, a linearly polarized planewave as an incident light is a good approximation[37] for estimating $F_z$. We calculate electromagnetic field distributions at various $r_{nb}$, $\theta$ and $\varphi$ using the finite element method (FEM) and then calculate $F_z$ using the Maxwell stress tensor[20,27]. More details are provided in the Method section.

We find certain geometrical windows that induce either positive or negative $F_z$ (Figs. 2b and 2c). For $r_{nb}$ > 90 nm (Fig. 2b) and $\theta$ < 75° (Fig. 2c), negative $F_z$ can be achieved, where the nano-bubble locates at the back side of the NP with respect to the incident light direction ($k_z$). Otherwise, $F_z$ is positive. $F_z$ is found insensitive to the azimuthal angle $\varphi$ (Fig. 2c). While the positive $F_z$ is intuitive, we need to understand the negative $F_z$ on the NP under illumination of the single planewave. Since $F_z$ is related to the electromagnetic energy density profile around the NP, we investigate the normalized absolute electric field profiles ($|E|/|E_0|$) for representative cases. Without the nano-bubble, $F_z$ on the NP have the same directions as the incident wavevector ($k_z$), and the corresponding $|E|/|E_0|$ profiles around the NP are spreading towards the $F_z$ directions (Figs. 2d or 2e). For the case where the NP experiences a negative force due to the presence of a nano-bubble (Fig. 2f), we find parts of the field profile, especially the strong field portion (pink region), spreads towards the -$k_z$ direction (the $F_z$ direction), which resembles that of the bare NP under the negative $k_z$ illumination (Fig. 2d). When the NP is close to the right surface of the bubble (Fig. 2g), the force is positive, and the strong field profile is similar to that of the bare NP under the positive $k_z$ illumination (Fig. 2e).

Figure 3a shows the calculated $F_z$ on a supercavitating NP along the central axis of the Gaussian beam for two representative cases: one for the positive motion ($r_{nb}$=130 nm and $\theta = 180°$) and one for the negative motion ($r_{nb}$=130 nm and $\theta = 0°$). The NP is found to experience $F_z$ on the order of ~$10^{-12}$ N around the focal plane (Fig. 3a), with the negative $F_z$ uniformly smaller than the positive $F_z$ in amplitude. It is noted that the profiles of the optical forces, which correspond to the light intensity profile, confirm that the optical gradient forces are negligible, which would have changed the sign of the force across the focal plane. Meanwhile, the Stokes' friction force in a flow with a low Reynolds number ($Re$), which is $Re = $ ~$10^{-3}$ <<1 in our case, allows us to estimate the order of magnitude of the driving force ($F_{\text{driving}}$) on the ballistic NP given their measured speeds as: $F_{\text{driving}} = 6\pi\eta r v \approx 6\pi \times 8.9\times10^{-4}$ (kg m$^{-1}$s$^{-1}$, viscosity of liquid water) $\times 6\times10^{-8}$ (m, radius of Au NP) $\times 1\times10^{-1}$ (m s$^{-1}$, speed of ballistic NP) = ~$10^{-10}$ N. This is two orders of magnitude larger than our calculated $F_z$ since we used liquid water viscosity. As the plasmonic Au NP is intensely excited by the laser to instantaneously evaporate the surrounding water and generate a nano-bubble to encapsulate itself, *i.e.*, supercavitation, the dynamic viscosity experienced by the NP should be different. Using the calculated optical force (~$10^{-12}$ N) and Stokes' law, we can extract an effective dynamic viscosity ($\eta_{eff}$) experienced by the ballistic NPs as:

$$\eta_{eff} = \frac{(z_2 - z_1)}{6\pi r \langle v \rangle} \left[ \int_{z_1}^{z_2} \frac{1}{F_z(z)} dz \right]^{-1} \tag{1}$$

where $\langle v \rangle$ is the experimentally determined average speed between positions $z_1$ and $z_2$ where the ballistic movement occurs. The averaged $\eta_{eff}$ (for the cases in Figs. 1d and 1e) is ~$1.5\times10^{-5}$ kg m$^{-1}$s$^{-1}$ for the negative motion and ~$1.1\times10^{-5}$ kg m$^{-1}$s$^{-1}$ for the positive motion. These values are much smaller than the dynamic viscosity of liquid water but very close to that of vapor steam (~$1\times10^{-5}$ kg m$^{-1}$s$^{-1}$), suggesting that the ballistic NPs are moving in a gaseous environment. This can be interpreted as the following: as the laser-excited Au NP encapsulated by a bubble moves forward, it keeps evaporating water, maintaining a vapor cushion in front of it and extends the bubble boundary forward (Fig. 3b). This is very similar to the observed near-zero drag force on a hot metal sphere enclosed by a gas cavity due to the Leidenfrost effect[38] and in great analogy to the supercavitating torpedoes, which realized 5× faster speed than conventional ones[39]. As the NP moves, the trailing end of the bubble, which becomes further away from the hot NP, cools and vapor condenses back to liquid (Fig. 3b). This interpretation is well backed by the results of our molecular dynamics (MD) simulations. In Figs. 3c, we clearly see that a hot NP encapsulated by a vapor bubble can continuously evaporate liquid molecules in front of it as it moves forward with a speed of ~13 m s$^{-1}$, so that the bubble boundary is extended and the NP is always enclosed in a gaseous environment. While this simulation uses a Lennard-Jones argon model system, the physics should be the same and analogy can be drawn with our NP-in-water experiment. We note that the 13 m s$^{-1}$ speed simulated is much higher than our experimental observation (0.1-0.3 m s$^{-1}$), but the NP is still found able to instantaneously extend the bubble front while moving.

Since nano-bubble dynamics is stochastic, not all Au NPs are enclosed by nano-bubbles to exhibit the ballistic movement. We can see that the ballistic motion generally occurs within 300 μm at either side of the focal plane with a significant portion of the NPs in this region exhibiting ballistic motion. It is possible that in this region the laser intensity is larger and can generate bubbles more easily. In addition, Figs. 2b and 2c show that the geometrical window for the negative $F_z$ is narrower than that for the positive $F_z$, which may explain our observation that more ballistic NPs move positively than negatively. The magnitude of the positive $F_z$ being generally larger than the negative $F_z$ (Fig. 3a) also explains the observed higher average speed of the positive motion than that of the negative motion (Fig. 1). Although $F_z$ depends on the instantaneous geometrical configuration of the NP-bubble structure and thus the forces should be dynamic, we observe that the NPs move almost strictly along the beam axis (Figs. 1d and 1e). Moving in a virtually frictionless supercavitation, the ballistic NPs exhibit speed-to-body length ratios greater than 10$^6$. This is at least 5 orders of magnitude higher than other directed nano/micro swimmers and 1-6 orders of magnitude higher than random micro swimmers (Fig. 4).

In theory, any nanoscale objects that can have supercavitation after laser excitation should exhibit similar ballistic movements. To confirm that this effect is generalizable, we have performed experiments with Au nanorods, which also have a SPR peak of 800 nm. It is observed that Au nanorods also exhibit similar ballistic movements with an observed speed of ~13,000 μm s$^{-1}$. Since the SPR characteristic of Au nanorod is anisotropic due to its shape, only the nanorods that align with the incident electric field direction can experience strong SPR coupling and lead to supercavitation. Thus, the ballistic movements of Au nanorods may easily be interrupted, because the different forces (*e.g.*, optical force and drag force) can make nanorods to misalign with the electric field. However, in Figure S4b, we can clearly see a glowing dot moving for ~ 13 μm within 1 ms along the beam propagating direction. This ballistic Au nanorod shows a normalized speed of 260,000 body-length s$^{-1}$, which is still ~$10^2$-$10^5$ times faster than the reported nano/micro swimmers driven by optical forces[17,18]. We have not observed any negative motion for Au nanorods, most likely due to the stringent alignment requirements that limit the probability of enabling supercavitation and negative optical forces. Nevertheless, this result demonstrates that the ballistic movement can be potentially generalized to other kinds of plasmonic NPs with different composition, geometry and dimensions. We believe that with proper NP design, both the ballistic motion and optical force direction can be manipulated and optimized.

**Materials and Methods**

Sample preparation: A quartz cuvette (Hellma, Sigma-Aldrich, 10 mm × 10 mm) with 4 windows is pre-cleaned with acetone, isopropyl alcohol and deionized water in an ultrasonic bath. A silica-Au core-shell NP-water suspension (Auroshell, Nanospectra Biosciences, Inc) with a NP number density of $2×10^{15}$ # m$^{-3}$ or an Au nanorod NP-water suspension (NanoXact, nanoComposix) with a NP number density of $2.6×10^{17}$ # m$^{-3}$ is mixed with deionized water to achieve different number densities.

Characterization of ballistic Au NPs: Mode-locked femtosecond laser pulses (center wavelength of 800 nm and a full-width half maximum of 10 nm) are emitted from a Ti:sapphire crystal in an optical cavity (Spectra Physics, Tsunami). The laser is collimated and a linearly polarized TEM$_{00}$ mode beam. The time interval between pulses is 12.4 ns (80.7 MHz), and the pulse duration is ~200 fs. The power of the laser is fixed to 690 mW. The femtosecond laser passes through a 20× objective lens and focuses on the NP-water suspension in the cuvette. A beam profiler (Thorlab BP104-UV) measures the minimum waist of the Gaussian beam (~ 6 μm). We note ballistic movements were still observed with a 10× objective lens or a laser power as high as ~1 W. The scattered lights from the Au NPs passes through another objective lens and focuses on the image sensors of the high-speed camera (HX-7, NAC). The recorded images in the camera are then analyzed using a customized image processing software in MATLAB.

Calculation of optical force: The finite element method (FEM) software (RF module, COMSOL Multiphysics) is used to calculate the optical forces on different Au NPs with nano-bubbles (structures illustrated in Fig. 2a). The Au NP consists of a silica core (radius of 50 nm) and an Au shell (thickness of 10 nm). It is modeled that an Au NP with a nano-bubble is immersed in an infinite volume of water. In the simulations, for the infinite volume of water, a spherical volume of water (radius of 400 nm) is enclosed by a perfectly matched laser with a thickness of 130 nm. The Au NP with a nano-bubble is located at the center of the water sphere. The refractive indices of water, air, and silica at the wavelength of 800 nm are 1.33, 1.00, and 1.50, respectively. The real and complex parts of the refractive index of the Au shell at the wavelength of 800 nm are 0.154 and 4.91, respectively. A linearly polarized planewave is modeled as the incident light for Figs. 2b, 2c, and 2d. A Gaussian beam given by the paraxial approximation[37] is used for Fig. 3a and extracting $\eta_{eff}$, where it has a minimum waist of 6 μm and the maximum light intensity of 12 mW μm$^{-2}$, corresponding to the experimental conditions. After solving the electromagnetic field distribution of the simulation domain, optical force ($\vec{F}$) is calculated as:

$$\vec{F} = \oiint \overleftrightarrow{T} \cdot \hat{n} \, \mathrm{d}a$$

where $\overleftrightarrow{T}$ is the time-averaged Maxwell stress tensor[20,27], and $\hat{n}$ is the normal vector of the surface of the Au NP. We use the time-averaged Maxwell stress tensor to calculate $\vec{F}$ in the z-direction on the surface of the Au NP.

**References**

1. Strebhardt, K. & Ullrich, A. Poul Ehrlich's magic bullet concept: 100 years of progress. *Nat. Rev. Cancer* **8**, 473-480 (2008).

2. Sanchez, M. M. & Schmidt, O. G. Medical microbots need better imaging and control. *Nature* **545**, 406-408 (2017).

3. Luo, M., Feng, T., Wang, T. & Guan, J. Micro-/nanobots at work in active drug delivery. *Adv. Funct. Mater.* **28**, 1706100 (2018).

4. Gao, W., Sattayasmitsathit, S. & Wang, J. Catalytically propelled micro-/nanomotors: how fast can they move?. *Chem. Rec.* **12**, 224-231 (2012).

5. Manjare, M., Yang, B. & Zhao, Y.-P. Bubble driven quasioscillatory translational motion of catalytic micromotors. *Phy. Rev. Letts* **109**, 128305 (2012).

6. Baylis, J. R. et al. Self-propelled particles that transport cargo through flowing blood and halt hemorrhage. *Sci. Adv.* **1 (9)**, e1500379 (2015).

7. Kagan, D. et al. Acoustic droplet vaporization and propulsion of perfluorocarbon-loaded microbullets for targeted tissue penetration and deformation. *Angew. Chem. Int. Ed.* **51**, 7519-7522 (2012).

8. Purcell, E. M. Life at low Reynolds number. *Am. J. Phys.* **45**, 3-11 (1977).

9. Sing, C. E., Schmid, L., Schneider, M. F., Franke, T. & Kats, A. A. Controlled surface-induced flows from the motion of self-assembled colloidal walkers. *Proc. Natl. Acad. Sci.* **107(2)**, 535-540 (2010).

10. Ghosh, A. & Fischer, P. Controlled propulsion of artificial magnetic nanostructured propellers. *Nano Letts.* **9**, 2243-2245 (2009).

11. Wu, Z. et al. Turning erythrocytes into functional micromotors. *ACS Nano* **8 (12)**, 12401-8 (2014).

12. Xuan, M. et al. Near infrared light-powered janus mesoporous silica nanoparticle motors. *J. Am. Chem. Soc.* **138 (20)**, 6492-6497 (2016).

13. Xuan, M. et al. Self-propelled nanomotors for thermomechanically percolating cell membranes. *Angew. Chem. Int. Ed.* **57**, 12463-12467 (2018).

14. Chen, C. et al. Light-steered isotropic semiconductor micromotors. *Adv. Mater.* **29**, 1603374 (2017).

15. Wang, Y. et al. Bipolar electrochemical mechanism for the propulsion of catalytic nanomotors in hydrogen peroxide solutions. *Langmuir* **22(25)**, 10451-10456 (2006).

16. Ibele, M., Mallouk, T. E. & Sen, A. Schooling behavior of light-powered autonomous micromotors in water. *Angew. Chem. Int. Ed.* **48**, 3308-3312 (2009).

17. Konlger, A. & Kohler, W. Optical funnerling and trapping of gold colloids in convergent laser beam. *ASC Nano* **6**, 4400-4409 (2012).

18. Kajorndejnukul, V., Ding, W., Sukhov, S., Qiu, C.-Q., Dogariu, A. Linear momentum increase and negative optical forces at dielectric interface. *Nat. Photon.* **7**, 787-790 (2013)
11

19. Ashkin, A., Dziedzic, J. M., Bjorkholm, J. E. & Chu, S. Observation of a single-beam gradient force optical trap for dielectric particles. *Opt. Letts.* **11**, 288-290 (1986).

20. Lehmuskero, A., Johansson, P., Dunlop, H. R., Tong, L. & Kall, M. Laser trapping of colloidal metal nanoparticles. *ACS Nano* **9**, 3453-3469 (2015).

21. Dogariu, A., Sukhov, S. & Saenz, J. J. Optically induced 'negative forces'. *Nat. Photon.* **7**, 24-27 (2013).

22. Chen, J., Ng, J., Lin, Z. & Chan, C. T. Optical pulling force. *Nat. Photon.* **5**, 531-534 (2011).

23. Bzrobohaty, O. et al. Experimental demonstration of optical transport, sorting and self-arrangement using a 'tractor beam'. *Nat. Photon.* **7**, 123-127 (2013).

24. Damkova, J. et al. Enhancement of the 'tractor-beam' pulling force on an optically bound structure. *Light Sci. Appl.* **7**, 17135 (2018).

25. Gao, D., Shi, R., Huang, Y. & Gao, L. Fano-enhanced pulling and pushing optical force on active plasmonic particles. *Phys. Rev. A* **96**, 043826 (2017).

26. Mizrahi, A. & Fainman, Y. Negative radiation pressure on gain medium structures. *Opt. Letts.* **35**, 3405-3407 (2010).

27. Guo, G., Feng, T. & Xu, Y. Tunable optical pulling force mediated by resonant electromagnetic coupling. *Opt. Letts.* **43**, 4961-4964 (2018).

28. Xu, L., Mou, F., Gong, H., Luo, M. & Guan, J. Light-driven micro/nanomotors: from fundamentals to applications. *Chem. Soc. Rev.* **46**, 6905-6926 (2017).

29. Seol, Y., Carpenter, A. E. & Perkins, T. T. Gold nanoparticles: enhanced optical trapping and sensitivity coupled with significant heating. *Opt. Letts.* **31**, 2429-2431 (2006).

30. Fu, X., Chen, B., Tang, J. & Zewail, A. H. Photoinduced nanobubble-driven superfast diffusion of nanoparticles imaged by 4D electron microscopy. *Sci. Adv.* **3**, e1701160 (2017).

31. Lukianova-Hleb, E., Volkov, A. N. & Lapotko, D. O. Laser pulse duration is critical for generation of plasmonic nanobubbles. *Langmuir* **30**, 7425-7434 (2014).

32. Boulais, E., Lachaine, R. & Meunier, M. Plasma mediated off-resonance plasmonic enhanced ultrafast laser-induced nanocavitation. *Nano Lett.* **12**, 4763-4769 (2012).

33. Metwally, K., Mensah, S. & Baffou, G. Fluence threshold for photothermal bubble generation using plasmonic nanoparticles. *J. Phys. Chem. C* **119**, 28586-28596 (2015).

34. Sasikumar, K. & Keblinski, P. Molecular dynamics investigation of nanoscale cavitation dynamics. *J. Chem. Phys.* **141**, 234508 (2014).

35. Merabia, S., Keblinski, P., Joly, L., Lewis, L. J. & Barrat, J.-L. Critical heat flux around strongly heated nanoparticles. *Phys. Rev. E* **79**, 021404 (2009).

36. Maheshwari, S., van der Hoef, M., Prosperetti & A., Lohse, D. Dynamics of formation of a vapor nanobubble around a heated nanoparticle. *J. Phys. Chem. C* **122**, 20571-20580 (2018).

37. Trojek, J., Chvatal, L. & Zemanek, P. Optical alignment and confinement of an ellipsoidal nanorod in optical tweezers: a theoretical study. *J. Opt. Soc. Am. A* **29**, 1224-1236 (2012).

38. Vakarelski, I. U. et al. Self-determined shapes and velocities of giant near-zero drage gas cavities. *Sci. Adv.* **3**, e1701558 (2017).

39. Ng, K. Overview of the ONR supercavitating high-speed bodies program. *AIAA Guidance, Navigation, and Control Conference and Exhibit* (2006).

**Author contributions**

All authors designed the experiments. E.L. set up and performed the experiments, and performed the optical simulations. D. H performed the MD simulations. All authors discussed the results and the mechanism of the ballistic Au NPs and wrote the manuscript. The authors declare no competing financial interest.

**Acknowledgements**

This work is supported by National Science Foundation and the Center for the Advancement of Science in Space.

# Figures

# Figure 1

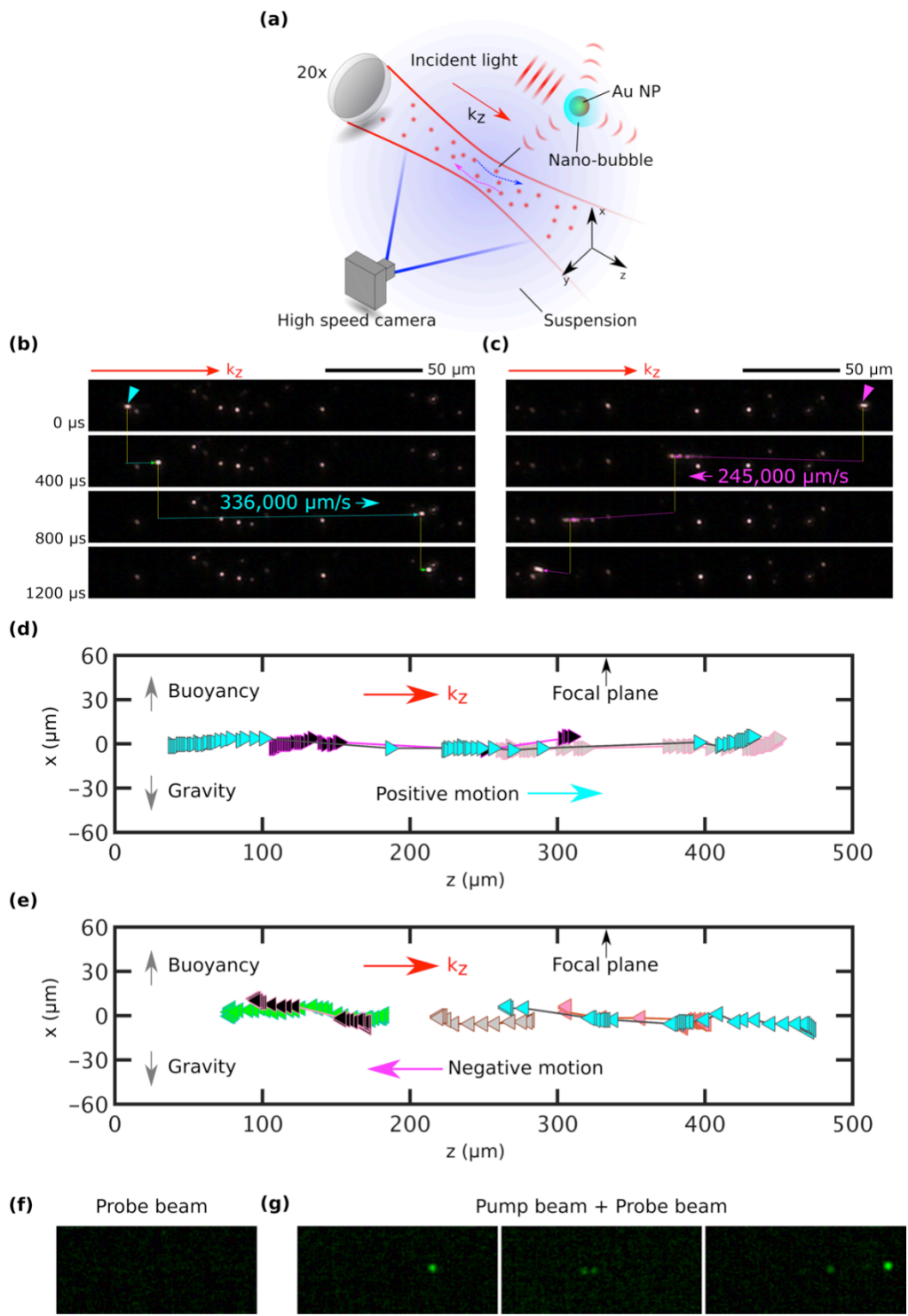

**Figure 1| Observation of extremely fast ballistic movements of Au NPs** (**a**) Schematic of the experimental system to monitor the dynamics of Au NPs dispersed in water. The red solid lines after the 20× lens indicate the Gaussian beam envelope. Au NPs scatter the incident light (red dots), which is captured by a high-speed camera with the field of view depicted by the blue solid lines. The blue and magenta arrows depict the Au NPs moving in the +z and -z directions, respectively referred to as the "positive motion" and the "negative motion". (**b** and **c**) Dark-field optical images (scale bars – 50 μm) of ultra-fast ballistic Au NPs with (b) the positive motion and (c) the negative motion. The cyan triangle in (b) and the magenta triangle in (c) indicate the positions of the selected Au NPs at t = 0. The cyan dotted line in (b) and the magenta dotted line in (c) are guide for the eye. The power of the laser is 690 mW, corresponding to 12 mW μm$^{-2}$ at the focal plane. (**d** and **e**) Representative (d) positive and (e) negative motion trajectories of the ballistic Au NPs projected onto the x-z plane. Each color corresponds to a ballistic Au NPs. The time interval between two adjacent symbols is 400 μs. The black arrows at the top of the figures indicate the focal plane location. In (a-d), $k_z$ is the wavevector of the incident laser light. (**f** and **g**) Images from the pump-probe optical scattering imaging experiment for (f) without pump beam and (g) with pump beam. Each green spot corresponds to the diffraction-limited scattered probe light from the Au NPs with nanobubbles.

**Figure 2**

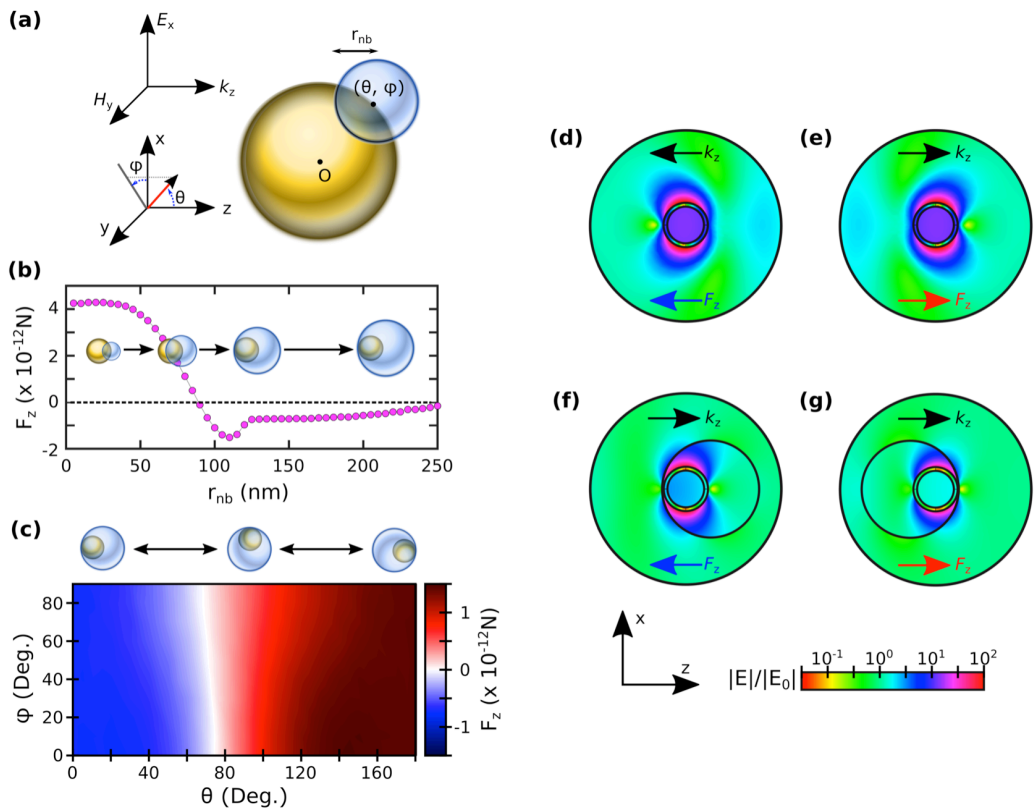

**Figure 2| Optical force on an Au NP with a nano-bubble** (**a**) Schematic of a model of an Au NP with a nano-bubble. The Au NP (yellow sphere) consists of a 100 nm-diameter silica core and a 10 nm-thick Au shell. The nano-bubble with a radius of $r_{nb}$ is attached to the surface $(\theta, \varphi)$ of the Au NP, where $\theta$ and $\varphi$

are, respectively, the polar and the azimuthal angles in the polar coordinate with an origin ($o$) at the center of the Au NP. $E_x$, $H_y$, and $k_z$ depict the electric field, the magnetic field, and the wavevector of the incident planewave, respectively. (**b**) The calculated optical force in the z-direction ($F_z$) as a function of $r_{nb}$. Here, $\theta = 0°$, $\varphi = 0°$ and the amplitude of $E_x$ is $2.6 \times 10^6$ V m$^{-1}$. The insets illustrate the schematic configurations of Au NP with a nano-bubble with different $r_{nb}$. (**c**) The calculated $F_z$ as a function of $\theta$ and $\varphi$. Here, $r_{nb}$ = 130 nm and the amplitude of $E_x$ is $2.6 \times 10^6$ V m$^{-1}$. On top of the contour, schematic configurations of Au NP with a nano-bubble as a function of $\theta$ are illustrated. (**d-g**) The calculated absolute electric field ($|E|$) profile normalized by the amplitude of incident electric field ($|E_0|$) for (d) a bare Au NP with the negative $k_z$, (e) a bare Au NP with the positive $k_z$, (f) an Au NP with a nano-bubble ($r_{nb}$ = 130 nm and $\theta = 0°$) and the positive $k_z$, and (g) an Au NP with a nano-bubble ($r_{nb}$ = 130 nm and $\theta = 180°$) and the positive $k_z$. Directions of $F_z$ are also shown.

**Figure 3**

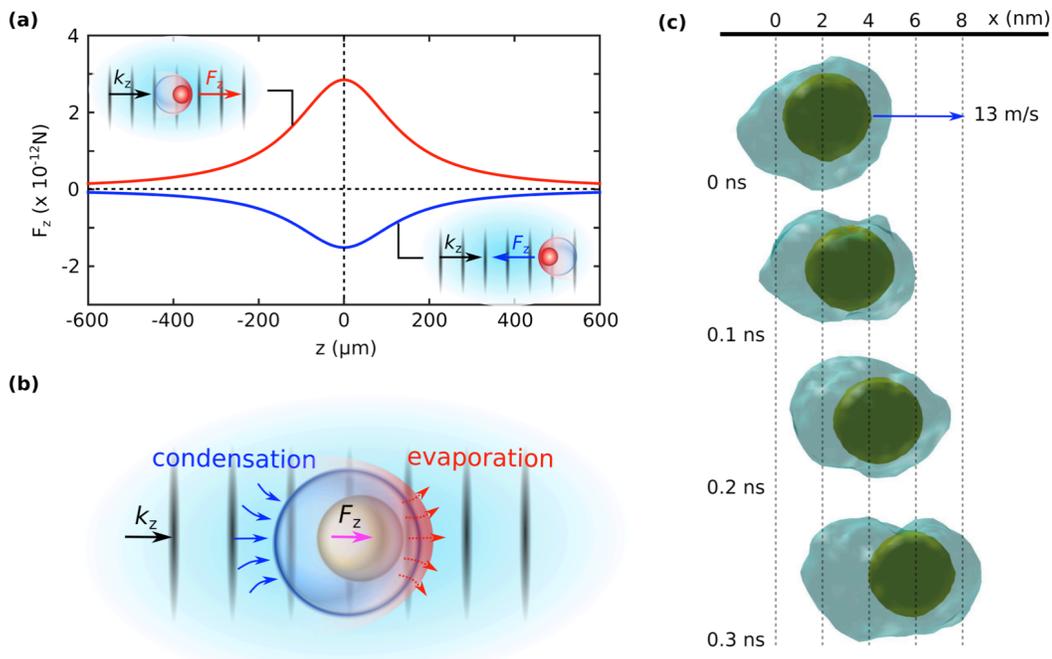

**Figure 3| Positive and negative optical forces and the nanoscale Leidenfrost effect** (**a**) The calculated $F_z$ on an Au NP with a nano-bubble along the central axis (z-direction) of the Gaussian beam. A beam waist and intensity at the focal plane ($z = 0$) are 6 μm and 12 mW μm$^{-2}$, respectively. The red line represents the positive force for $r_{nb}$ = 130 nm, $\theta = 180°$, and the blue line represents the negative force $r_{nb}$ = 130 nm, $\theta = 0°$. The insets illustrate (left top) the positive motion and (right bottom) the negative motion. (**b**) Schematic illustration of a supercavitating ballistic Au NP. Under the laser illumination, the plasmonic Au NP is intensely heated to generate a nano-bubble which encapsulates it. When the Au NP moves forward, it keeps evaporating water and maintaining a vapor cushion in front of it, which effectively makes the NP moving in a virtually frictionless environment, like the Leidenfrost effect. In the meantime, vapor at the trailing end of the bubble cools and condenses back to liquid as the hot NP moves forward. (**c**) Nanoscale Leidenfrost effect of a hot moving NP simulated by molecular dynamic (MD) simulations. The hot NP is thermostated at 1000 K, and the blue contour visualizes the isosurface of the critical density of liquid argon (0.536 g cm$^{-3}$), which represents the nanobubble surface. The NP moves in the positive x-direction with a speed of 13 m s$^{-1}$.

**Figure 4**

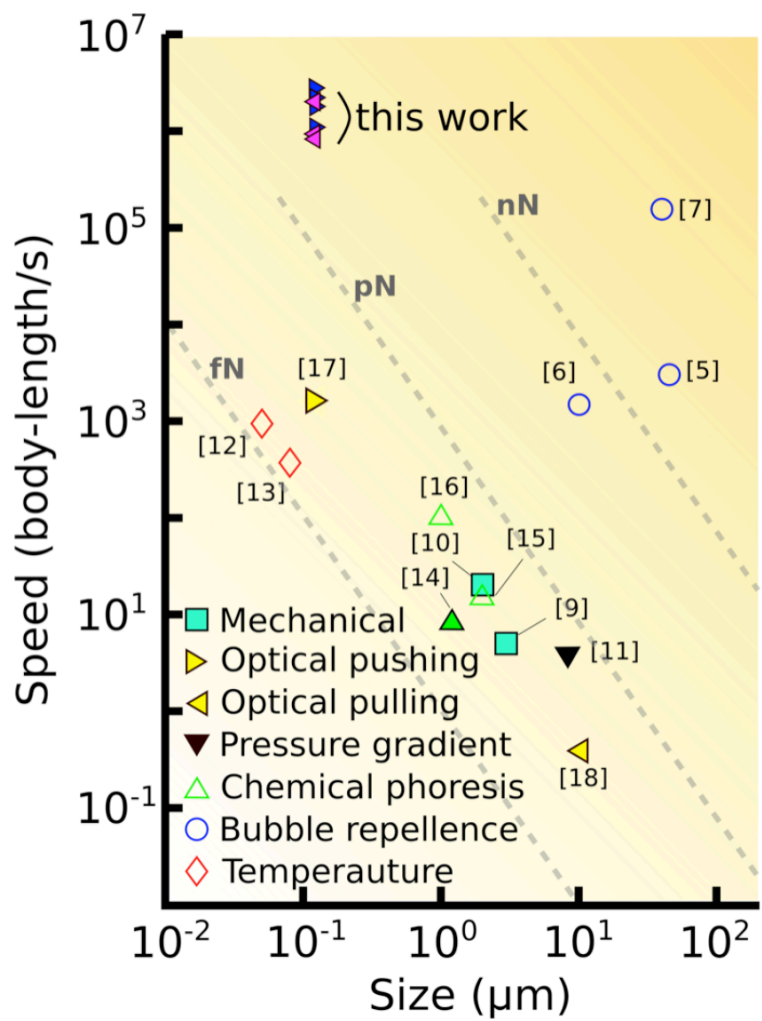

**Figure 4| Comparison of the ballistic Au NP swimmers with other swimmers** Top-four highest speeds (measured in body-length s$^{-1}$) for the negative and the positive motion compared to those of the nano/micro swimmers in the literature. The filled symbols represent nano/micro swimmers whose movement can be directed. The hollow symbols represent those moving in random directions. The legend indicates propulsion mechanisms. The gray dash lines are contour lines of the magnitude of the force needed to achieve the corresponding speed of a swimmer in liquid water.

17